\documentclass{article}

\usepackage{epsfig,amsfonts,amssymb,newlfont,rotate,float,array}

\begin{document}

\title{Critical exponents of the three dimensional\\
diluted Ising model}

\author{
H.~G.~Ballesteros,
L.~A.~Fern\'andez,\\
V.~Mart\'{\i}n-Mayor,
A.~Mu\~noz Sudupe,\\
{\it Departamento de F\'{\i}sica Te\'orica I,}\\ 
{\it Universidad Complutense de Madrid,}\\
{\it 28040 Madrid, Spain,}\\
{\tt\small  hector, laf, victor, sudupe@lattice.fis.ucm.es}
\\
\\
G.~Parisi and
J.~J.~Ruiz-Lorenzo \\
{\it Dipartimento di Fisica and INFN,}\\
{\it  Universit\`a di Roma I, La Sapienza,}\\ 
{\it P. A. Moro 2, 00185 Roma, Italy,}\\
{\tt\small   giorgio.parisi@roma1.infn.it, ruiz@chimera.roma1.infn.it}
}
\date{February 25, 1998}

\maketitle

\begin{abstract}
We study the phase diagram of the site-diluted Ising model in a wide
dilution range, through Monte Carlo simulations and Finite-Size
Scaling techniques. Our results for the critical exponents and
universal cumulants turn out to be dilution-independent, but only after a
proper infinite volume extrapolation, taking into account the leading
corrections-to-scaling terms. 
\end{abstract}

\bigskip


\newpage

\section{Introduction}

The magnetic phase diagram and critical properties of many
magnetic materials can be described by a simple model (the
Heisenberg Hamiltonian):
\begin{equation}
H=\sum_{i,j,\alpha,\beta} J^{ij}_{\alpha\beta}\, S_i^{\alpha}S_j^{\beta}.
\label{ACCIONGEN}
\end{equation}
Here $S_i^{\alpha}$ is a spin operator (Latin indices refer to lattice
sites, while Greek ones represent the spin components). 
$J^{i j}_{\alpha \beta}$ is a coupling matrix which
is usually short-ranged and can represent either the Ising, XY or
Heisenberg models by properly dealing with the spin index.  One can
understand eq.~(\ref{ACCIONGEN}) on the basis of the exchange
interaction between the electrons of the external shells of the
atoms. This interaction, expressed as in eq.~(\ref{ACCIONGEN}), is
symmetric in the spin index. Nonetheless if one puts the atoms on a
crystalline lattice, the material tends to magnetize in the so-called
axes or planes of easy magnetization given by the symmetry of the
crystal.

One typical example are the uniaxial crystals, as the hexagonal
lattices, where the magnetization can choose as
subspace of easy magnetization the $c$ axis or its orthogonal
plane. In the first case the system is well described assuming that
the magnetic momenta point in the $c$ direction and it should be
described by the Ising model.  In the second one, the material should be
studied by means of the XY model.

However no pure material exists in Nature, and  it is mandatory
to consider the effects of non-magnetic impurities. The simplest way
to do so, is by considering a modified version of (\ref{ACCIONGEN})
\begin{equation}
H=\sum_{i,j,\alpha,\beta} J^{ij}_{\alpha\beta}\,\epsilon_i\,\epsilon_j S_i^{\alpha}S_j^{\beta}
 ,
\label{ACCIONGENQUEN}
\end{equation}
where the $\epsilon$'s are quenched, uncorrelated random variables,
chosen to be 1 with probability $p$ (the spin concentration) or 0 with
probability $1-p$ (the impurity concentration, or spin dilution). 
The rationale for the quenched approximation is that usual
relaxation times for the non-magnetic impurities are much longer than
the corresponding for spin dynamics. For non-frustrated systems, the phase diagram of
(\ref{ACCIONGENQUEN}) in the temperature-dilution plane consists of a
magnetically disordered (paramagnetic) region at high temperature, 
separated from an ordered (ferromagnetic) region  at lower temperatures. 
The dilution dependent critical
temperature, $T_{\mathrm c}(p)$, obviously equals the pure model
value at $p=1$. It lowers for larger dilution
values, until the extreme case $T_{\mathrm c}(p_{\mathrm c})=0$ at the
site percolation threshold for the concentration of
the magnetic atoms. 

Not many general results have been obtained for
the Hamiltonian (\ref{ACCIONGENQUEN}). The most popular one is
doubtless the {\it Harris criterion}~\cite{HARRIS}. 
This criterion states that the critical behavior of
(\ref{ACCIONGENQUEN}) will be the same as for (\ref{ACCIONGEN}) if the
specific-heat critical exponent, $\alpha$, is negative, while a new
Universality class will appear if $\alpha>0$.  In the latter case its
is possible to show \cite{CHAYES} that $\alpha$ for the diluted model 
is negative:
one of the effects of the dilution is to smooth the critical behavior
of the system.  The only one between the generic models for magnetism
(Ising, XY, Heisenberg) displaying $\alpha>0$ in three dimensions, is
the Ising model.

There are other physical contexts in which the Hamiltonian 
(\ref{ACCIONGENQUEN}) has been studied. For instance, its
four dimensional Ising version has been recently investigated (see
\cite{ISDIL4D} and references therein) in connection with
the puzzling problem of finding non asymptotically-free interacting
theories in four dimensions. The two dimensional model is
also interesting as a playground for exactly solvable field-theories,
and has also been considered (see \cite{ISDIL2D,ZEROES,2Dbis} and references 
therein).

As already stated, the materials displaying Ising-like behavior in very
pure samples should behave differently when the impurities concentration
increases. In fact, according to Harris, an infinitesimal impurity 
concentration should be enough to spoil the Ising behavior. However
this will happen in very narrow intervals of  temperatures, which may
be unreachable experimentally. 

The Hamiltonian (\ref{ACCIONGENQUEN}) can be studied in the low dilution
regime by means of analytical perturbative renormalization-group
methods~\cite{NEWMAN,JUG,MAYER}.  They find a new fixed-point, thus
implying that the critical exponents along the $T_{\mathrm c}(p)$ line
are dilution independent and different from their pure Ising
value. The predicted correlation length exponent, $\nu$,
ranges from 0.697 in ref.~\cite{NEWMAN} to 0.67 in
ref.~\cite{MAYER}. This should be compared with the Ising model
result, $\nu=0.6300(15)$, given in ref.~\cite{ZINN-JUSTIN}.  For the order
parameter critical exponent, $\beta$, a value of 0.35 is obtained,
contrasting with 0.3250(15) for pure Ising~\cite{ZINN-JUSTIN}. For the
susceptibility exponent, the perturbative analysis predicts
$\gamma=1.32$ and the corresponding Ising one~\cite{ZINN-JUSTIN} is
1.241(2). This is maybe the quantity more easily comparable with
experimental results, as the magnetic susceptibility can be very
precisely measured.

The study of the Hamiltonian (\ref{ACCIONGENQUEN}) beyond the low
disorder regime, is restricted to the Monte Carlo (MC) method.  Many
simulations have been performed in the last seventeen years
\cite{LANDAU,MARROOTHERS,WANG,HOLEY,HENNECKE,HEUER}.  The first study,
on small lattices~\cite{LANDAU} was compatible with the new fixed
point scenario. However further simulations \cite{MARROOTHERS} found
results rather suggesting a continuously varying value of the critical
exponents along the critical line. A Monte Carlo Renormalization Group
study~\cite{HOLEY} found a value for the $\nu$ exponent consistent
with the perturbative one at $p=0.8$. However, for $p=0.9$ their
results did not differ from the pure Ising model, while for $p<0.8$
they could not find meaningful results. More recent
simulations~\cite{WANG} suggested a single fixed point scenario with
$\nu=0.77(4)$, confirmed in ref.~\cite{HENNECKE} where $\nu=0.78(1)$
was found at $p=0.4$. This puzzle of mutually contradicting results
started to make sense in ref~\cite{HEUER}. In this work, the crucial
observation that the exponents measured in a finite lattice are
transitory was made. Unfortunately the statistical errors at large
dilution did not allow for a definite conclusion.

When writing this paper a new MC work on this model has appeared
\cite{WISEMAN}. They obtain $\nu=0.682(2)$ at $p=0.8$ but a
markedly different result ($\nu=0.717(8)$) at $p=0.6$.

In this paper we present the first sound numerical evidence for a
random fixed point dominant along the whole critical line. 
This is achieved by means of
a Finite-Size Scaling (FSS) analysis, in a wide dilution range 
($0.4\leq p\leq 0.9$, the percolation threshold being at 
$p_{\mathrm c} \approx 0.31$~\cite{STAUFFER}). 
The investigation of very diluted
samples is made possible by a $p$-reweighting  method, which allows
to extrapolate the simulation results obtained at $p$ to a close $p'$
value~\cite{PERC,ISDIL4D,ISDIL2D}. 
A careful consideration of the
scaling corrections is needed, in order to get the right value in the
infinite volume limit.  In this system, the first
corrections-to-scaling exponent, $\omega$, is very small
($\omega\approx0.4$, see ref.~\cite{NEWMAN}).  Thus, the confusing
results in previous MC studies can be understood as an unusually
large contribution of the scaling corrections. After a proper
consideration of this problem, we find dilution independent 
critical exponents in quantitative 
agreement with perturbative calculations. 

Other theoretical problem of interest is the absence of {\it
self-averaging} at the critical point. This means that the
disorder-realization variance of quantities such as the magnetic
susceptibility or the specific-heat, at the critical point, is a fixed,
non-zero fraction of their mean values even in the thermodynamical limit. It
has been argued~\cite{AHARONY} that this fixed fraction is an
universal number. In ref.~\cite{ISDIL4D}, this fraction for the
susceptibility is calculated analytically and numerically in four
dimensions. In this work, we numerically calculate this ratio, along
the critical line $T_{\mathrm c}(p)$. After the compulsory
infinite volume extrapolation, an universal, dilution independent
result is found. A very recent simulation~\cite{WISEMAN} has questioned the
universality of these ratios. However, these
authors do not perform any infinite volume extrapolation, making their
conclusions necessarily not definitive.

The experimental study is still not completed. For instance,
indications of the expected new universality class were obtained in the Ising
antiferromagnet Fe$_{1-x}$Zn$_x$F$_2$, studied in the reduced
temperature range $10^{-3}\leq t \leq 10^{-1}$ \cite{BARRET,BIRGENEAU}. In this
system the order parameter exponent was found to be $\beta=0.36$
\cite{BARRET}, while the obtained susceptibility exponent, $\gamma$,
was 1.44(6)\cite{BIRGENEAU}. Also in ref.~\cite{BIRGENEAU}, a
cusp-like behavior of the specific-heat was found, so no divergence
was expected.  This yields $\nu\geq 2/3$ through standard
hyperscaling relations. Another system investigated was
a dysprosium aluminum garnet doped with yttrium~\cite{HASTINGS},
for which $\beta=0.385(25)$ was obtained at a $5\%$ dilution.
The results regarding the $\beta$ exponent
have been questioned in ref.~\cite{THURSTON} where
Mn$_{0.5}$Zn$_{0.5}$F$_2$ was studied by synchrotron magnetic X-rays
scattering. These authors conclude that the experimental errors to date
are too big to distinguish between the pure Ising and the diluted
$\beta$ values. Maybe the strongest evidence found for a new
Universality Class has been reported in ref.~\cite{MITCHELL} studying
Mn$_{1-x}$Zn$_x$F$_2$ by means of neutron scattering.  Critical
exponents $\nu=0.70(2)$ and $\gamma=1.37(4)$ were found.

The layout of the paper is as follows. In section 2 we 
define the model and the observables to be measured in the numerical 
simulation. In section 3 we provide the necessary technical
details about the MC methods. Section 4 is devoted to Finite Size Scaling
techniques. After that, in section 5, we present our numerical results
and discuss the need for an infinite-volume extrapolation. This is
considered in section 6. We present our conclusions in section 7.

\section{The Model}
We have considered the site-diluted Ising model on the
single-cubic lattice, with nearest neighbors interaction.
We will work in a lattice of linear size $L$, with periodic boundary 
conditions.
The Hamiltonian is
\begin{equation}
H=-\beta\sum_{<i,j>} \epsilon_i \epsilon_j \sigma_i \sigma_j\ ,
\label{HAMILTONIANO}
\end{equation}
where $\sigma$ are the usual Z$_2$ spin variables. The $\epsilon$'s
are the quenched random variables introduced in (\ref{ACCIONGENQUEN}).  We
shall refer to an actual $\{\epsilon_i\}$ configuration as a {\it
sample}. We study the so-called quenched disorder: that is,
for every observable it is understood that we {\it first} calculate
the average on the $\{\sigma_i\}$ variables with the Boltzmann weight
given by  $\exp (-H)$, the results on the different samples
being {\it later} averaged. 

To avoid confusions, we will denote the Ising average with brackets,
while the subsequent sample average will be overlined. The observables will be
denoted with calligraphic letters, i.e.  $\cal O$, and with italics the
double average $O=\overline{\langle\cal O\rangle}$.
The total nearest-neighbor energy is defined
as 

\begin{equation}
{\cal E} =\sum_{\langle
i,j\rangle}\epsilon_i\sigma_i\epsilon_j\sigma_j\ .
\end{equation}
The energy is extensively used for extrapolating the results obtained for an
observable, $O$, at coupling $\beta$ to a nearby $\beta'$ 
coupling \cite{REWEIGHT} 
and for calculating $\beta$-derivatives through its connected correlation.
For instance, one can define the specific-heat as
\begin{equation}
C \, =  \partial_\beta \overline{<\cal E>} \,=
\frac{1}{V}\left( \, \overline{\langle\cal E^2\rangle-{\langle\cal E\rangle}^2} \, \right),
\end{equation}
$V$ being the total number of sites in the lattice, $L^3$.

The normalized magnetization is 
\begin{equation}
{\cal M}=\frac{1}{V}\sum_i \epsilon_i\sigma_i\ .
\end{equation}
In terms of the magnetization we can give a convenient definition of the 
susceptibility as 
\begin{equation}
\chi=V\overline{\left\langle {\cal M}^2 \right\rangle}\ ,
\end{equation}
its Binder parameter being
\begin{equation}
g_4=\frac{3}{2}-\frac{1}{2}\frac{\overline{\langle {\cal M}^4\rangle}}
           {\overline{\langle {\cal M}^2 \rangle}^2}\ .
\end{equation}
Another kind of cumulant, meaningless for the pure system, can be
defined as
\begin{equation}
g_2=\frac{\overline{ \langle {\cal M}^2 \rangle^2 - 
\overline{\langle {\cal M}^2 \rangle}^2
}}
{\overline{ \langle {\cal M}^2 \rangle}^2 } \ .
\label{g2}
\end{equation}
This quantity would be zero in the thermodynamical limit if self-averaging
is to be found. 
A very useful definition of the correlation length in a finite lattice, 
reads~\cite{XIL}

\begin{equation}
\xi=\left(\frac{\chi/F-1}{4\sin^2(\pi/L)}\right)^\frac{1}{2},
\label{XI}
\end{equation}
where $F$ is defined in terms of the Fourier transform of the
magnetization
\begin{equation}
{\cal G}(\mbox{\boldmath$k$})=
\frac{1}{V}\sum_{\mbox{\boldmath\scriptsize$r$}}{\mathrm e}^{\mathrm i
\mbox{\boldmath\scriptsize$k$}\cdot\mbox{\boldmath\scriptsize$r$}}
\epsilon_{\mbox{\boldmath\scriptsize$r$}}
\sigma_{\mbox{\boldmath\scriptsize$r$}}\ ,
\end{equation}
as
\begin{equation}
F=\frac{V}{3}\overline{\left\langle |{\cal
G}(2\pi/L,0,0)|^2+\mathrm{permutations}\right\rangle}\ .
\end{equation}

This definition is very well behaved for the FSS method 
we employ~\cite{OURFSS}. 

\section{The Monte Carlo Update}

The method of choice for an Ising model simulation is a
cluster-method~\cite{CLUSTER}. The most efficient variety for the pure
model is the Wolff single-cluster update~\cite{WOLFF}.  However, in
diluted systems, very small (even individual) groups of nearly (or
completely) isolated spins can appear. These groups are scarcely
changed with a single cluster method. Thus we have constructed our
elementary MC step (EMCS) as 250 cluster flips complemented with a
Metropolis step. For the largest dilutions ($p=0.4, 0.5$) the presence
of isolated intermediate-sized groups of spins makes the
thermalization too slow.  For these dilutions, in the EMCS we have
carried out a standard Swendsen-Wang sweep every $200$ single-cluster
flips.  We discard 100 EMCS for equilibration, then measuring after
every EMCS. The autocorrelation times for all observables are very
small (near 1 EMCS in the largest lattice), 
but we have also controlled that our update method
correctly thermalizes, by comparing hot and cold starts.

A disordered model simulation gets characterized by two parameters,
the number of samples generated ($N_S$), and the number of independent 
measures taken in each sample ($N_I$). Previous works 
(for instance~\cite{HOLEY,HENNECKE,HEUER}) have
chosen the \hbox{$N_I\gg N_S$} regime. However (see~\cite{ISDIL4D}),
the optimal regime is 
\begin{equation}
N_I\sim\left(\frac{\sigma_I}{\sigma_S}\right)^2,
\label{NI}
\end{equation}
where $\sigma_I$ is the mean variance {\it in a sample} of the
observable under consideration, while $\sigma_S$ is the variance {\it
between different samples}. Moreover, the non-vanishing value of $g_2$
shows that the susceptibility is not a self-averaging quantity, thus
making very dangerous the small $N_S$ regime. In this
work we have fixed \hbox{$N_I=200$} and $N_S=20000$. For $p=0.9$ we
performed $N_S=10000$.

In addition to the usual $\beta$ extrapolation~\cite{REWEIGHT}, in
some cases it is useful to perform a $p$ extrapolation. It can be done
as we know the precise distribution of the densities of the actual
configurations (binomial distribution). Details of the method can be
found in ref.~\cite{ISDIL4D} for the same model in four dimensions.

We remark that the large number of samples used, combined with the
relative small number of measures, makes the $\beta$ extrapolations
biased.  A proper statistical procedure allows to cancel the bias.  We
address to ref.~\cite{ISDIL4D} for details about the method we follow.

\section{Finite Size Scaling Methods}

A very efficient way of measuring critical exponents~\cite{OURFSS}
follows from this form of the FSS Ansatz
\begin{equation}
O(L,\beta,p)=L^{x_O/\nu}\left(F_O(\xi(L,\beta,p)/L)+O(L^{-\omega})\right)\
,
\label{FSS}
\end{equation}
where a critical behavior $t^{-x_O}$ is expected for the operator $O$
and $F_O$ is a (smooth) scaling function. 
From a Renormalization Group point of view, $\omega$ is
the eigenvalue corresponding to the leading irrelevant operator.
It is very important that, in the above
equation, only quantities measurable on a finite lattice appear.
Notice that terms of order $\xi^{-\omega}_{L=\infty}$ are dropped from
eq.~(\ref{FSS}), so we assume that we are deep within the scaling
region. 

To eliminate the unknown scaling function, we measure
the quotient 
\begin{equation}
Q_O=O(sL,\beta,p)/O(L,\beta,p)\ ,
\end{equation}
at the coupling value for which the correlation length in units of
the lattice size is the same for both lattices. So we get
\begin{equation}
\left.Q_O\right|_{Q_\xi=s}=s^{x_O/\nu}+O(L^{-\omega})\ .
\label{QUO}
\end{equation}
Given the strong statistical correlation between $Q_O$ and $Q_\xi$,
the above quotient can be obtained with great accuracy (in fact,
in our opinion, this is the best method available to measure the 
usually tiny three-dimensional $\eta$ exponents~\cite{OURFSS}). 

In many cases (high precision computations or small lattices), it is
useful to parameterize the leading corrections-to-scaling, 
thus we need to
consider in the analysis a behavior like 
\begin{equation}
\left.Q_O\right|_{Q_\xi=s}=s^{x_O/\nu}+A^O_p \, L^{-\omega} + \cdots
\label{QUOMEGA}\ .
\end{equation}
Here the dots stand for higher-order corrections, while $A^O_p$ is a
dilution-dependent slope. 

The most convenient observables to measure the two independent
critical exponents, $\eta$ and $\nu$, are found to be
\begin{eqnarray*}
\partial_\beta\xi&\rightarrow& x=\nu+1\ ,\\
\chi&\rightarrow& x=\nu(2-\eta).
\end{eqnarray*}

\section{Numerical results}

\begin{figure}[t]
\begin{center}
\leavevmode \epsfig{file=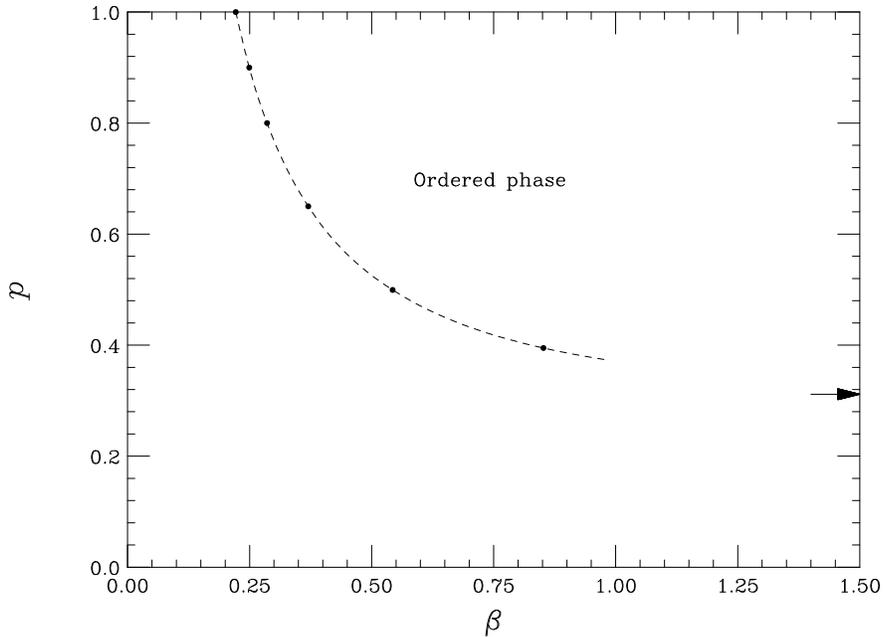,width=0.68\linewidth,angle=90}
\end{center}
\caption{Phase diagram of the model (\ref{HAMILTONIANO}), in the
inverse temperature--dilution plane. The dots correspond to the simulated
points, while the arrow signals the percolation
limit ($\beta=\infty$).}
\label{DIAGRAMA}
\end{figure}

The phase diagram of the model (\ref{HAMILTONIANO}) is shown
in figure~\ref{DIAGRAMA}. In this work
we have simulated  lattices
$L=8,16,32,64$ and $128$, at dilutions $p=0.9,0.8,0.65,0.5$ and
$0.4$. Our procedure has been the following. For $p=0.9,0.8,0.65$, we
have chosen a $\beta$ coupling value where the relation
\begin{equation}
\frac{\xi(L,\beta,p)}{L}=\frac{\xi(2L,\beta,p)}{2L}
\label{MATCHING}
\end{equation} 
approximately holds. Then, we have relied on standard reweighting
methods, which allow to extrapolate the simulation results at coupling
$\beta$ to a close $\beta'$, to precisely fulfill the matching
condition~(\ref{MATCHING}). For very diluted systems, 
the transition line is almost horizontal (see figure~\ref{DIAGRAMA}) 
thus it is more
convenient to use a $p$-reweighting method to extrapolate the
simulation results to a nearby $p'$ value (see
refs.~\cite{PERC,ISDIL4D,ISDIL2D}). Therefore, we have first located
the $\beta$ values for which eq.~(\ref{MATCHING}) holds at
$p=0.4,0.5$, then we have fixed this $\beta$ value, and changed $p$
later on.  In this way, the true critical dilutions for fixed $\beta$,
$p_{\mathrm c}(\beta)$, differ from $0.4$ and $0.5$ (in less than a
$2\%$).  Nevertheless, we shall keep referring to them as $p=0.4,0.5$
in tables and graphics, for the sake of clarity.

In table~\ref{TABLERAW} we present the results for exponents $\nu$ and
$\eta$ and cumulants $g_4$ and $g_2$, using eq.~(\ref{QUO})
(neglecting scaling corrections). Beware that consecutive data in the
table are anticorrelated (the results of lattice $L$ are used once in
the numerator and another time in the denominator in
eq.~(\ref{QUO})). For the error computation we have used a jack-knife
method with 50 bins, ensuring a 10\% of uncertainty in the error
bars. Thus, we display 2 digits in these bars if the first one is smaller
than 5.

Notice that the exponent $\eta$ and the cumulant $g_4$ are, before any
infinite volume extrapolation, quite dilution independent. This can be
understood because they show a very mild evolution with the lattice
size. On the contrary, exponent $\nu$ and cumulant $g_2$ show a larger
dependence on the lattice size and so, an infinite volume
extrapolation is needed before one can extract definite
conclusions. Nevertheless, one can already guess from the table that $\nu$
is surely different from the pure Ising value and the $g_2$ cumulant
is different from zero (there is not self-averaging). The latter was
also observed in the same model in four dimensions \cite{ISDIL4D},
where we found mean field results plus logarithmic corrections.

\begin{table}[t]
\caption{Critical quantities obtained from pairs $(L,2L)$ using 
eq.~(\ref{QUO}) for all the dilutions simulated.}
\begin{center}
\smallskip
\begin{tabular*}{\hsize}{@{\extracolsep{\fill}}lrlllll}\hline
& 
\multicolumn{1}{l}{$L$}  & 
\multicolumn{1}{c}{$p=0.9$} &
\multicolumn{1}{c}{$p=0.8$} &
\multicolumn{1}{c}{$p=0.65$} &
\multicolumn{1}{c}{$p=0.5$} &
\multicolumn{1}{c}{$p=0.4$}\\\hline\hline
$\eta$&8&.0171(7)&.0219(7)&.0284(10)&.0296(24)&.0322(29)\\
     &16&.0277(7)&.0308(7)&.0330(8) &.0345(19)&.0297(16)\\
     &32&.0320(9)&.0335(8)&.0329(9) &.0313(11)&.0315(17)\\
     &64&.0349(9)&.0346(8)&.0335(8) &.0329(14)&.0326(13) \\\hline

$\nu$&8 &.6663(14)&.6877(11)&.7172(16)&.7447(24)&.7718(32)\\
     &16&.6643(14)&.6849(12)&.7107(18)&.7328(22)&.7534(32)\\
     &32&.6631(15)&.6836(12)&.7048(20)&.7189(24)&.7382(27)\\
     &64&.6644(15)&.6864(14)&.6996(20)&.7118(21)&.7182(26) \\\hline

$g_2$&8 &.0832(10) &.1546(16)&.2310(25)&.2784(24)&.3043(24)\\
     &16&.0861(12) &.1500(14)&.2077(15)&.2371(20)&.2551(22)\\
     &32&.0918(13) &.1474(17)&.1920(20)&.2138(22)&.2296(25)\\
     &64&.0974(17) &.1477(12)&.1842(19)&.1994(21)&.2106(16)\\\hline

$g_4$&8 &.7049(14)&.6900(17)&.6814(23)&.6900(20)&.6989(20)\\
     &16&.6926(17)&.6818(15)&.6809(16)&.6871(18)&.6958(21)\\
     &32&.6876(19)&.6819(16)&.6832(20)&.6879(17)&.6889(20)\\
     &64&.6821(16)&.6771(18)&.6780(17)&.6825(19)&.6857(22)\\\hline

\end{tabular*}
\label{TABLERAW}
\end{center}
\end{table}

\begin{figure}[t]
\begin{center}
\leavevmode
\epsfig{file=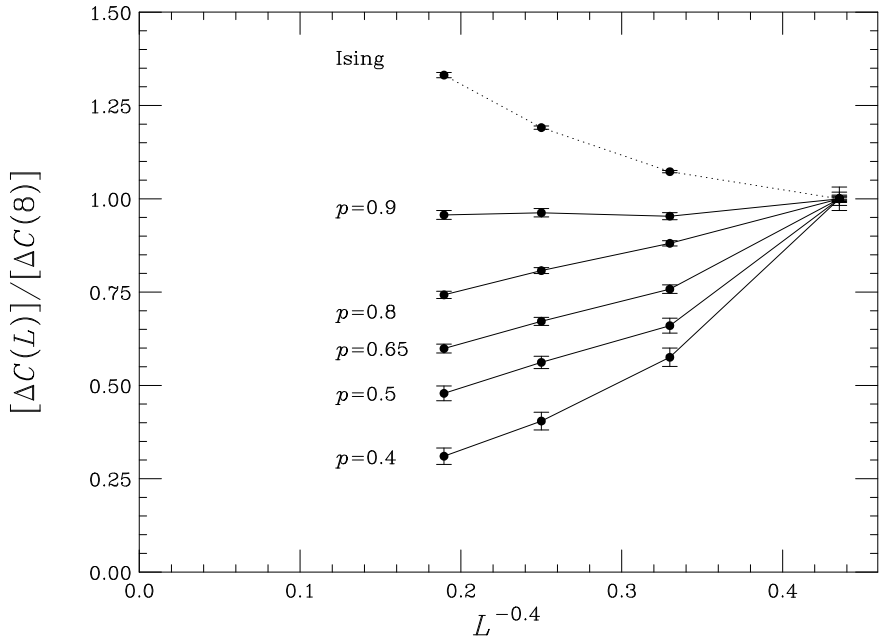,width=0.68\linewidth,angle=90}
\end{center}
\caption{Normalized specific-heat difference at the point where $Q_\xi=2$. 
The $\omega\approx0.4$ value used in the plot is obtained in 
ref.~\cite{NEWMAN}.}
\label{DELTAC}
\end{figure}

Another quantity of interest is the specific-heat.  As stated in the
introduction, $\alpha$ is negative and no divergences are
expected. This is a quite difficult behavior to study, because FSS
investigations in other models displaying $\alpha<0$,
show that the specific-heat at the critical point is a growing, 
though bounded, quantity~\cite{OURFSS}.  For this reason we choose to study
$$\Delta C(L)= \left[C(2L)-C(L)\right] _{Q_\xi=2}.$$ This quantity
diverges if $\alpha>0$, tends to zero if $\alpha<0$ and goes to a
constant value if the specific-heat diverges logarithmically
($\alpha=0$).  In addition, the (usually large) background term of the
specific-heat disappears. It will be convenient  to
recall that deriving the FSS Ansatz from the Renormalization 
Group~\cite{BARBER}, one finds a behavior for the specific-heat as $L^{2
y_T-d}$ (where $y_T=1/\nu$). Therefore one should expect the
fulfillment of hyperscaling relations for the transient exponents,
$\alpha(L)$ and $\nu(L)$.
In figure~\ref{DELTAC} we plot the $\Delta
C(L)$ values obtained.
As a contrast we also plot the corresponding
values for the pure Ising model which grow, as they should (the data
are taken from ref.~\cite{ISING}).  
We find a
decreasing value of $\Delta C(L)$ for $p\leq 0.8$, as expected.
Notice that the (transient) $\nu\approx 2/3$ found for $p=0.9$ in 
table~\ref{TABLERAW}, implies $\alpha=0$ through hyperscaling 
relations. This is very nicely shown in the plot,
where a constant value of $\Delta C(L, p=0.9)$ is seen.
Plotting $\Delta C(L)$ against $L^{\alpha/\nu}$ would be useless, because
the scaling corrections go approximately as $L^{-0.4}$, that is, their
lattice size evolution is much faster than that of the asymptotic
term.

\section{Infinite volume extrapolation}

As shown in the previous section, with our statistical accuracy the
values for the critical exponents are seen to depend on the lattice
size, so an infinite volume extrapolation is required (see
eq.~(\ref{QUOMEGA})).  However, one has to decide when the dots in
eq.~(\ref{QUOMEGA}) can be neglected. Our criterium will be the
following.  We perform the fit for lattice sizes not smaller than a given
$L_\mathrm{min}$.  If the fit quality is reasonable (i.e.  a not
too large $\chi^2/\mathrm{d.o.f.}$ calculated with the full
covariance matrix), we repeat it for lattices not smaller than
$2L_\mathrm{min}$. If this last fit is also reasonable and {\em the
extrapolated values are compatible in both fits}, we keep the central
value from the $L_\mathrm{min}$ fit, but quote error bars from the
$2L_\mathrm{min}$ one.

Therefore we need an estimate for $\omega$. 
We shall obtain it from the lattice size
evolution of the scaling functions:

\begin{eqnarray}
\left.\frac{\xi}{L}\right|_{Q_\xi=2}&=&{\left(\frac{\xi}{L}\right)}^\infty 
+ A_p^\xi L^{-\omega}+...\, ,\nonumber\\
\label{q1q2q3}
\vphantom{\left.\frac{\xi}{L}\right|_{Q_\xi=2}}
\left.g_4\right|_{Q_\xi=2}&=&g_4^\infty + A_p^{g_4} L^{-\omega}+...\, ,\\
\vphantom{\left.\frac{\xi}{L}\right|_{Q_\xi=2}}
\left.g_2\right|_{Q_\xi=2}&=&g_2^\infty + A_p^{g_2} L^{-\omega}+...\, 
.\nonumber
\end{eqnarray}

Then we shall use this $\omega$ value to extrapolate the critical exponents
$\nu$ and $\eta$. A reasonable value of $\chi^2/\mathrm{d.o.f.}$ 
in these fits will be a consistency condition.
A technical point of interest is that the  single universality-class scenario
requires the infinite volume extrapolation for $s^{x_O/\nu}$ to be
dilution-independent. Therefore, we can include data of different
dilutions and lattice sizes in the fit. 

\begin{figure}[t]
\begin{center}
\leavevmode
\epsfig{file=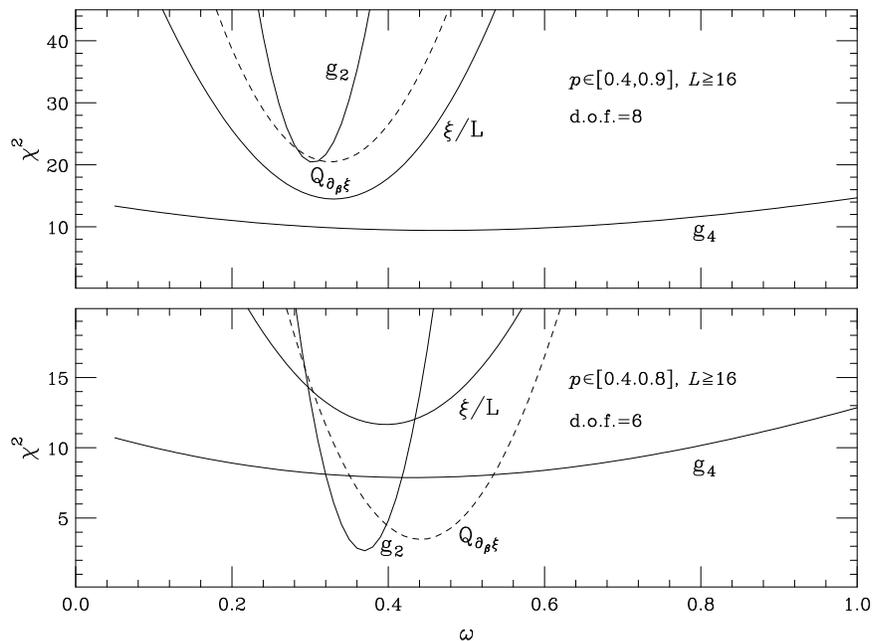,width=0.68\linewidth,angle=90}
\end{center}
\caption{Minimum of $\chi^2$ as a function of $\omega$, for the fits of
eq.~(\ref{q1q2q3}). We also plot with a dashed line the corresponding
quantity for the $Q_{\partial_\beta \xi}$ fit.
}
\label{CHI_OMEGA}
\end{figure}

In fig. \ref{CHI_OMEGA}, we plot the minimum of
$\chi^2/\mathrm{d.o.f.}$ in a fit to eq.~(\ref{q1q2q3}), as a
function of $\omega$. Several points become clear. It is obvious that
$g_4$ is not useful at all in order to fix $\omega$ (this is not
surprising as it shows almost no scaling corrections, $A_p^{g_4}\approx 0$). 
We see that including the $p=0.9$ data yields
an untenable fit with $L_\mathrm{min}=16$. Moreover, when we study the
extrapolation for $Q_{\partial_\beta\xi}=2^{1+1/\nu}$, we find an awful
result. This could have been anticipated from figure~\ref{DELTAC},
where a clearly non-asymptotic value for the specific-heat at $p=0.9$
is seen.  On the contrary, discarding the $p=0.9$ data, reasonable
fits are obtained.  Thus, we conclude that the $p=0.9$ system is still
crossing-over from the pure Ising fixed point to the diluted one, even
for lattices as large as $L=128$. Finally, it is evident from the plot
that the determination of $\omega$ can be greatly improved by means of
a {\it joint fit} of the $g_2$ and $\xi/L$ scaling functions.  The
results for this fit are shown in table~\ref{OMEGAJOINT}.
According to our, conservative, dots-neglecting criterium, we find
\begin{equation}
\label{OMEGA}
\omega=0.37(6).
\end{equation}
Notice that the value obtained in \cite{NEWMAN}, $\omega=0.42$ 
(without error estimation),
using the scaling-field method for momentum-space RG equations, is
compatible with ours.

\begin{table}[t]
\caption{Results of the infinite volume extrapolation of $g_2$
and $\xi/L$, including data from $L\geq L_\mathrm{min}$, at
$p=0.4,0.5,0.65$ and $0.8$. $Q(\chi^2,\mathrm{d.o.f.})$ is the
probability of getting a larger $\chi^2$ in the fit.
}
\smallskip 
\begin{center}
\begin{tabular*}{\linewidth}{@{\extracolsep{\fill}}rccccc}\hline
\multicolumn{1}
{c}{$L_\mathrm{min}$} 
&$\chi^2/{\mathrm{d.o.f.}}$&$Q$&{$\omega$} 
    &$\xi/L$
    &$g_2$ \\\hline\hline
8 &46.2/21&0.0012&0.430(15)&0.5890(17)&0.1458(17)\\
16&15.0/13&0.31  &0.37(2)  &0.598(4)&0.145(3)\\
32&1.95/5 &0.86  &0.38(6)  &0.587(7)&0.150(7)\\\hline
\end{tabular*}
\label{OMEGAJOINT}
\end{center}
\end{table}

In table~\ref{EXTRAEXPO} we present the infinite volume extrapolation for
$\nu$ and $\eta$ critical exponents and the $g_4$ cumulant. 
We see that $L_\mathrm{min}=16$ fulfills our dots-neglecting criterium
for $g_4$ and $\eta$. For $\nu$, $L_\mathrm{min}=8$ is found to be enough.
Our final values are
\begin{eqnarray}
\nu&=&0.6837(24)(29),\nonumber\\
\label{EXPOPREF}
\eta&=&0.0374(36)(9),\\
g_4&=&0.673(7)(2),\nonumber
\end{eqnarray}
where the first error is statistical while the second is due to the 
uncertainty in $\omega$. From (\ref{EXPOPREF}) we obtain
\begin{eqnarray}
\alpha&=&-0.051(7)(9),\nonumber\\
\beta&=&0.3546(18)(10),\\
\gamma&=&1.342(5)(5).\nonumber
\end{eqnarray}
For the computation of the statistical error in $\beta$ and $\gamma$ we 
take into account that the statistical correlation between 
$\nu$ and $\eta$ has turned out to be negligible.

\begin{table}[t]
\caption{Infinite volume extrapolation and fit qualities for the
critical exponents, including data from $L\geq L_\mathrm{min}$, at
$p=0.4,0.5,0.65$ and $0.8$ using eq.~(\ref{QUOMEGA}). The second error
is due to the indetermination in $\omega=0.37(6)$.}
\begin{center}
\smallskip
\begin{tabular*}{\linewidth}{@{\extracolsep{\fill}}lrlcc}\hline
 &$L_\mathrm{min}$ & Extrapolation & $\chi^2/{\mathrm {d.o.f.}}$  
    &$Q$\\\hline\hline
$\nu$ & 8&0.6837(10)(29)&14.0/11&0.24\\
      &16&0.6838(24)(33)&6.26/7&0.51\\
      &32&0.687(6)(2)&4.14/3&0.25\\\hline
$\eta$& 8&0.0419(8)(20)&96.4/11&$ < 10^{-15}$\\
      &16&0.0374(12)(9)&8.92/7&0.26\\
      &32&0.0374(36)(8)&0.18/3&0.98\\\hline
$g_4$ & 8&0.6726(21)(25)&31.5/11&.0001\\
      &16&0.6734(28)(21)&7.95/7&0.34\\
      &32&0.665(7)(3)&1.08/3&0.78\\\hline
\end{tabular*}
\label{EXTRAEXPO}
\end{center}
\end{table}

\begin{figure}[ht]
\begin{center}
\leavevmode
\epsfig{file=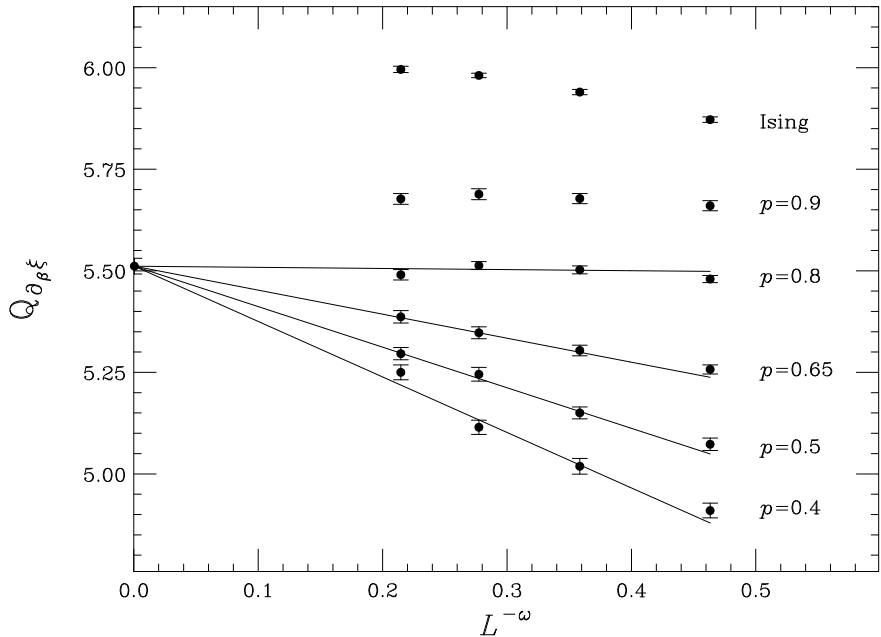,width=0.68\linewidth,angle=90}
\end{center}
\caption{$Q_{\partial_\beta\xi}=2^{1+1/\nu}$ 
for the different dilutions. The solid lines correspond to a fit enforced 
to yield the same infinite volume extrapolation for $p\leq 0.8$. 
The smallest lattice in the fit is $L=8$ and we use $\omega=0.37$. 
The Ising data have been taken from \cite{ISING}.
}
\label{NUDIL}
\end{figure}

\begin{figure}[ht]
\begin{center}
\leavevmode
\epsfig{file=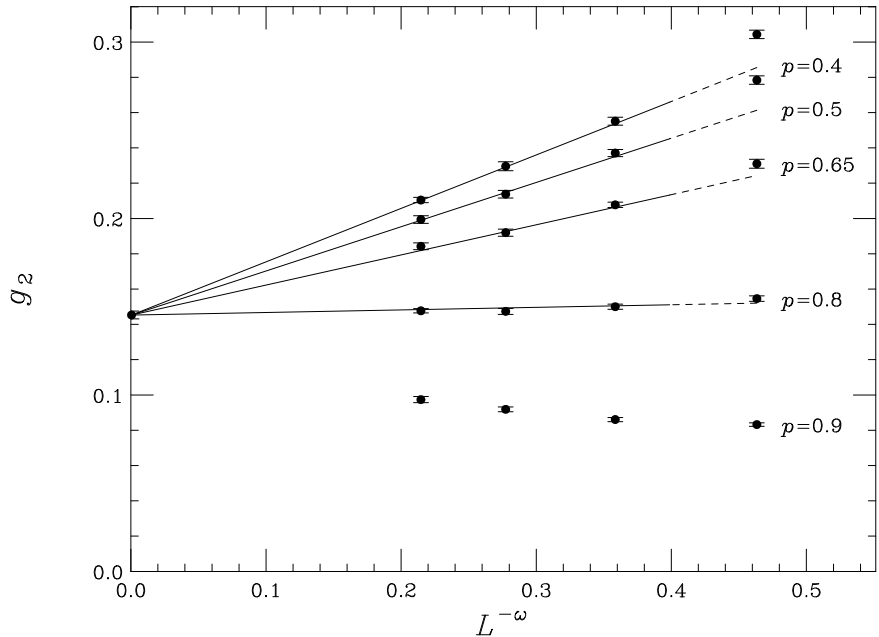,width=0.68\linewidth,angle=90}
\end{center}
\caption{Cumulant $g_2$ as a function of $L^{-\omega}$. 
The solid lines correspond to a fit enforced 
to yield the same infinite volume extrapolation for $p\leq 0.8$. 
The smallest lattice in the fit is $L=16$ and we use $\omega=0.37$.}
\label{G2}
\end{figure}

In figure~\ref{NUDIL} we show $Q_{\partial_\beta \xi}$ as a function
of $\omega$ for all the dilutions. We also plot the corresponding
values for the pure Ising model.  The solid lines correspond to the
joint fit for $L_\mathrm{min}=8$ using the data from $p\leq
0.8$.  Notice that the data are strongly
anticorrelated, therefore the apparent $\chi^2$ on the plot is larger
that the real one, computed with the full covariance matrix.  An
analogous fit for $g_2$ is shown in figure~\ref{G2}. We remark that the
$p=0.9$ data point to a maybe too low value. This is another signature of
the crossover from the Ising fixed point ($g_2=0$) to the diluted one.

\begin{table}[ht]
\caption{Crossing points of scaling functions $\xi/L$ and $g_4$ 
for pairs $L$ and $2L$ for the different dilutions.}
\begin{center}
\smallskip
\small
\begin{tabular*}{\linewidth}{@{\extracolsep{\fill}}crlllll}\hline
& 
\multicolumn{1}{c}{$L$}  & 
\multicolumn{1}{c}{$p=0.9$} &
\multicolumn{1}{c}{$p=0.8$} &
\multicolumn{1}{c}{$p=0.65$} &
\multicolumn{1}{c}{$p=0.5$} &
\multicolumn{1}{c}{$p=0.4$}\\\hline\hline
$g_4$&8 &.249583(30) &.286002(48) &.37025(13)  &.49996(27) &.39577(28)\\ 
     &16&.249340(15) &.285765(18) &.370185(36) &.49949(6)  &.39512(8)\\ 
     &32&.2492901(13)&.285758(7)  &.370208(16) &.499485(30)&.394895(33)\\ 
     &64&.2492924(15)&.2857417(25)&.3701649(48)&.499409(11)&.394840(13)\\ 
\hline                                                                     
$\xi/L$&8&.249299(26)  &.285690(49)&.36961(10)   &.49814(17) &.39302(19)\\  
       &16&.249291(12) &.285708(15)&.369986(31)  &.49896(5)  &.39441(6)\\ 
       &32&.2492957(44)&.285745(6) &.370147(13)  &.499326(21)&.394694(23)\\
       &64&.2492901(13)&.2857394(23)&.3701540(44)&.499374(9) &.394785(10)\\
\hline
\end{tabular*}
\label{CROSSINGRAW}
\end{center}
\end{table}

It is interesting to compare the  values for $g_4$ and $g_2$ with
those obtained in four dimensions~\cite{ISDIL4D}.
\begin{eqnarray}
g_4&=&0.32455,\nonumber\\
g_2&=&0.31024.\nonumber
\end{eqnarray}

Finally, we can compute the infinite volume critical couplings by
studying the crossing points of scaling functions (as $\xi/L$ and
$g_4$) measured in lattices of sizes $L$ and $sL$. Let $\Delta
\beta_\mathrm{c}^L$, $\Delta p_\mathrm{c}^L$ be the deviation of these crossing point from the
infinite-volume critical couplings.  The expected scaling behavior
is~\cite{BINDER}:
\begin{equation}
\Delta \beta_\mathrm{c}^L, \Delta p_\mathrm{c}^L
\propto \frac{1-s^{-\omega}}{s^{1/\nu}-1}L^{-\omega-1/\nu}\ .
\label{BETACFIT}
\end{equation}
In table~\ref{CROSSINGRAW} we present the crossing points of $\xi/L$
and $g_4$ for the $(L,2L)$ pair for all the dilutions simulated. We find
again that an infinite volume extrapolation is needed in order to
extract the critical couplings.

\begin{table}[ht]
\caption{Infinite volume critical couplings estimations for the
studied dilutions. The first error bar corresponds to the statistical
fit error, the second one (almost negligible) is due to the
uncertainty in $\omega+1/\nu$ exponent. For this table we use
$\omega+1/\nu=1.83(6)$.}
\begin{center}
\smallskip
\begin{tabular*}{\linewidth}{@{\extracolsep{\fill}}cllll}\hline
& 
\multicolumn{1}{c}{$L_{\mathrm {min}}$}  & 
\multicolumn{1}{c}{$\chi^2/{\mathrm {d.o.f.}}$} &
\multicolumn{1}{c}{$p_{\mathrm c}$}&
\multicolumn{1}{c}{$\beta_{\mathrm c}$}\\\hline\hline

&   16&   0.11/3 &0.394816(11)(2) &0.852 \\
&   32&   0.04/1 &  0.394821(22)(7)&0.852 \\\hline

&     16&   2.93/3 &  0.499413(9)(1) &0.543\\
&     32&   0.78/1 &  0.499394(17)(4)&0.543\\\hline
                                  
&      16&   5.27/3 &0.65 & 0.370166(5)(1)\\
&      32&   1.53/1 &0.65 & 0.370156(8)(0) \\\hline
                                  
&      16&   5.41/3 &0.8 &0.2857421(30)(0) \\
&      32&   0.27/1 &0.8 &0.2857368(47)(5)\\\hline
            
&      16&   8.45/3 &0.9 &0.2492905(19)(0) \\
&      32&   0.03/1 &0.9 & 0.2492880(30)(5)\\\hline
\end{tabular*}
\label{BETACRAW}
\end{center}
\end{table}

Using eq.~(\ref{BETACFIT}) for $s=2$ we perform a joint fit for both
scaling functions $g_4$ and $\xi/L$. For this fit we take 
$\omega+1/\nu=1.83(6)$. The final results for the different  dilutions
studied are shown in table~\ref{BETACRAW}, where two values for 
$L_\mathrm{min}$ are used. Let us remark that our critical couplings
are compatible with the results in~\cite{HEUER} 
($\beta_{\mathrm c}^{p=0.8}=0.28578(4)$, 
$\beta_{\mathrm c}^{p=0.9}=0.24933(3)$). But we definitely do not
agree with the value $\beta_{\mathrm c}^{p=0.8}=0.2857609(4)$ quoted
in ref.~\cite{WISEMAN}. This is not surprising 
as in this work the corrections-to-scaling are not considered.

\section{Conclusions}

We have shown, beyond the low-disorder limit, that the diluted
Ising model is in the basin of attraction of a single fixed point.  
Therefore, if randomness is to be modelized
with eq.(\ref{ACCIONGENQUEN}), the critical exponents of an Ising
system are not those of the pure Ising model, but those of the random
fixed-point (although this may be fairly hard to show in a very pure
sample). To establish this result we have simulated in a very wide
dilution range, finding a consistent picture {\it only after an
infinite volume extrapolation}.  The $p=0.9$ data seem, however, to be
still crossing-over from the pure Ising fixed-point to the diluted one
in lattices as large as $L=128$.

We obtain the values of the critical exponents and universal cumulants
eliminating the systematic errors coming from the leading
corrections-to-scaling terms. The previous computations did not consider
these terms and were not able to control the corresponding systematic
effects. Incidentally, most of the computations have been carried out
at $p=0.8$ as in this case the scaling corrections are very small, and
the results in small lattices seem stable. However, even in this case
the lack of an extrapolation produces an underestimation of the errors.

The (dilution-independent) critical exponents are shown to be in good
agreement with the series estimates~\cite{NEWMAN,JUG,MAYER}.  The
corrections-to-scaling exponent, $\omega$, is measured with a $16\%$
error and is found to be in quantitative agreement with the
perturbative estimate~\cite{NEWMAN}. The smallness of this exponent
explains why this problem is so hard to attack numerically. 
In fact, the total computer time devoted to this work has
been about 5 Intel Pentium-Pro years.  As we had already shown in four
dimensions~\cite{ISDIL4D}, diluted Ising models are found not to be
self-averaging at criticality in three dimensions (see
ref.~\cite{WISEMAN} for an independent verification in three
dimensions). This is proved by showing that the quotient between the
sample-variance of the susceptibility and its mean-value, tends in the
thermodynamic limit to a non-zero constant independent of the dilution
(it is a renormalization-group invariant). This quotient is measured
with a $4\%$ accuracy after the infinite volume extrapolation.

\section{Acknowledgments}

The computations have been carried out using the RTNN machines at Universidad
de Zaragoza and Universidad Complutense de Madrid. We acknowledge
CICyT for partial financial support (AEN97-1708 and AEN97-1693).
JJRL is granted by  EC HMC (ERBFMBICT950429).


\begin{thebibliography}{99}

\bibitem{HARRIS}
A. B. Harris, {\sl J. Phys.} {\bf C7}, 1671 (1974).

\bibitem{CHAYES}
J. T. Chayes, L. Chayes, D. S. Fisher and T. Spencer,
{\sl Phys. Rev. Lett} {\bf 57}, 2999 (1986).

\bibitem{ISDIL4D}
H. G. Ballesteros, L. A. Fern\'andez, V. Mart\'{\i}n-Mayor, 
A. Mu\~noz Sudupe, G.~Parisi and J.J. Ruiz-Lorenzo,
{\sl Nucl. Phys.} {\bf B512[FS]}, 681 (1998).

\bibitem{ISDIL2D}
H. G. Ballesteros, L. A. Fern\'andez, V. Mart\'{\i}n-Mayor, 
A. Mu\~noz Sudupe, G.~Parisi and  J. J. Ruiz-Lorenzo,
{\sl J. Phys.} {\bf A30} 8379 (1997).

\bibitem{ZEROES} J. J. Ruiz-Lorenzo, {\sl J. Phys.} {\bf A30}, 485 (1997).

\bibitem{2Dbis} F. D. A. Aar\~ao Reis, S. L. A. de Queiroz and R. R. dos 
Santos, {\sl Phys. Rev.} {\bf B56}, 6013 (1997).

\bibitem{NEWMAN}K. E. Newman and E. K. Riedel,
{\sl Phys. Rev. } {\bf B25},  264 (1982).

\bibitem{JUG} G. Jug, {\sl Phys. Rev. } {\bf B27},  609 (1983).

\bibitem{MAYER} I. O. Mayer, {\sl J. Phys. A: Math. Gen. }{\bf 22},
2815 (1989).

\bibitem{ZINN-JUSTIN}
J.C. Le Guillou and J. Zinn-Justin, {\sl Phys. Rev.} {\bf B21}, 3976 (1980).
 
\bibitem{LANDAU}D. P. Landau,
{\sl Phys. Rev.} {\bf B22},  2450 (1980).

\bibitem{MARROOTHERS} J. Marro, A. Labarta and J. Tejada,
{\sl Phys. Rev.} {\bf B34},  347 (1986);
D. Chowdhury and D. Stauffer,
{\sl J. Stat. Phys. } {\bf 44},   203 (1986);
P. Braun and M. F\"ahnle,
{\sl J. Stat. Phys. } {\bf 52},  775 (1988);
P. Braun {\it et al.},
{\sl Int. J. Mod. Phys. } {\bf B3}, 1343 (1989).

\bibitem{WANG} J. S. Wang and D. Chowdhury,
{\sl J. Phys. (Paris) } {\bf 50},  2905 (1989).

\bibitem{HOLEY} T. Holey and M. F\"ahnle,
{\sl Phys. Rev.} {\bf B41},  11709 (1990).

\bibitem{HENNECKE} M. Hennecke, {\sl Phys. Rev. } {\bf B48},  6271  (1993).

\bibitem{HEUER} H. O. Heuer, {\sl J. Phys. A: Math. Gen. }{\bf 26},
L333 (1993).

\bibitem{WISEMAN}
S. Wiseman and E. Domany, {\sl cond-mat}/9802095; {\sl cond-mat}/9802102.

\bibitem{STAUFFER} D. Stauffer and A. Aharony. {\em Introduction to the 
percolation theory}. (Taylor \& Francis, London 1994)

\bibitem{PERC}
H. G. Ballesteros, L. A. Fern\'andez, V. Mart\'{\i}n-Mayor, 
A. Mu\~noz Sudupe, G.~Parisi and J. J. Ruiz-Lorenzo,
{\sl Phys. Lett.} {\bf B400}, 346 (1997).

\bibitem{AHARONY} A. Aharony and A. B. Harris,
{\sl Phys. Rev. Lett. }{\bf 77}, 3700 (1996).

\bibitem{BARRET} P. H. Barret, {\sl Phys. Rev. } {\bf B34},   3513 (1986).

\bibitem{BIRGENEAU}R. J. Birgeneau {\it et al.},
{\sl Phys. Rev.} {\bf B27},  6747 (1983).

\bibitem{HASTINGS} J. M. Hastings, L. M. Corliss and W. Kunnmann,
{\sl Phys. Rev. } {\bf B31},  2902 (1985).

\bibitem{THURSTON}T. R. Thurston, C. J. Peters, R. J. Birgeneau, and
P. M. Horn, {\sl Phys. Rev. } {\bf B37},   9559 (1988).

\bibitem{MITCHELL} P. W. Mitchel {\it et al.},
{\sl Phys. Rev. } {\bf B34},   4719 (1986).


\bibitem{REWEIGHT}
M. Falcioni, E. Marinari, M. L. Paciello, G. Parisi and B. Taglienti,
{\sl Phys. Lett.} {\bf 108} 331 (1982) ;
A. M. Ferrenberg and R. H. Swendsen, {\sl
Phys. Rev. Lett.} {\bf 61}  2635 (1988).

\bibitem{XIL} F. Cooper, B. Freedman and D. Preston, 
{\sl Nucl. Phys. }{\bf B210}, 210 (1989).

\bibitem{OURFSS}
H. G. Ballesteros, L. A. Fern\'andez, V. Mart\'{\i}n-Mayor, and
A. Mu\~noz Sudupe,~{\sl Phys. Lett.}~{\bf B378} (1996) 207;
{\sl Phys. Lett.}~{\bf B387}, 125 (1996); {\sl Nucl. Phys.} {\bf B
483}, 707 (1997).

\bibitem{CLUSTER} R. H. Swendsen and J. S. Wang,
{\sl Phys. Rev. Lett.} {\bf 58}, 86 (1987).

\bibitem{WOLFF} U.~Wolff, {\sl Phys. Rev. Lett.} {\bf 62}, 3834 (1989).


\bibitem{BARBER} M. N. Barber, {\sl Finite-size Scaling } in 
{\it Phase Transitions and Critical phenomena}, edited by 
C. Domb and J.L. Lebowitz (Academic Press, New York, 1983) vol 8.

\bibitem{ISING}
H. G. Ballesteros {\it et al.}. Work in preparation.

\bibitem{BINDER} K. Binder, {\sl Z. Phys. } {\bf B43} 119 (1981).


\end{thebibliography}
\end{document}